\documentclass[journal=jpclcd,manuscript=article,layout=twocolumn]{achemso}

\usepackage{chemformula} 
\usepackage[T1]{fontenc} 
\mciteErrorOnUnknownfalse
\SectionNumbersOn

\usepackage{graphicx}
\usepackage{tabularx}
\usepackage{dcolumn}
\usepackage{bm}
\usepackage{color}
\usepackage[inkscapearea=page]{svg}
\usepackage{adjustbox}
\usepackage{float}

\usepackage{amsmath}
\usepackage{amssymb}
\usepackage{multirow}
\usepackage{booktabs}
\usepackage{caption}
\usepackage{subcaption}

\usepackage{soul}

\newcommand{\R}{\mathbb{R}}
\newcommand{\bigO}{\mathcal{O}}

\DeclareMathOperator{\JVP}{JVP}
\DeclareMathOperator{\VJP}{VJP}
\DeclareMathOperator{\Grad}{Grad}
\DeclareMathOperator{\Jacobian}{Jac}
\DeclareMathOperator{\Hess}{Hess}
\DeclareMathOperator{\trace}{trace}
\DeclareMathOperator{\Divergence}{Divergence}
\DeclareMathOperator{\Laplacian}{Laplacian}
\DeclareMathOperator{\HVP}{HVP}

\author{Niklas Frederik Schmitz}
\affiliation{Machine Learning Group, Technische Universität Berlin, 10587 Berlin, Germany}

\author{Klaus-Robert M\"uller}
\affiliation{Machine Learning Group, Technische Universität Berlin, 10587 Berlin, Germany}
\alsoaffiliation{BIFOLD - Berlin Institute for the Foundations of Learning and Data, 10587 Berlin, Germany}
\alsoaffiliation{Department of Artificial Intelligence, Korea University, Seongbuk-gu, Seoul 02841, Korea}
\alsoaffiliation{Max Planck Institute for Informatics, Stuhlsatzenhausweg, 66123 Saarbrücken, Germany}
\alsoaffiliation{Google Research, Brain Team, 10117 Berlin, Germany}
\email{klaus-robert.mueller@tu-berlin.de}

\author{Stefan Chmiela}
\affiliation{Machine Learning Group, Technische Universität Berlin, 10587 Berlin, Germany}
\alsoaffiliation{BIFOLD - Berlin Institute for the Foundations of Learning and Data, 10587 Berlin, Germany}
\email{stefan@chmiela.com}

\title
  {Algorithmic Differentiation for Automated Modeling of Machine Learned Force Fields}

\abbreviations{IR,NMR,UV}

\begin{document}

\begin{tocentry}

\includegraphics[width=\textwidth]{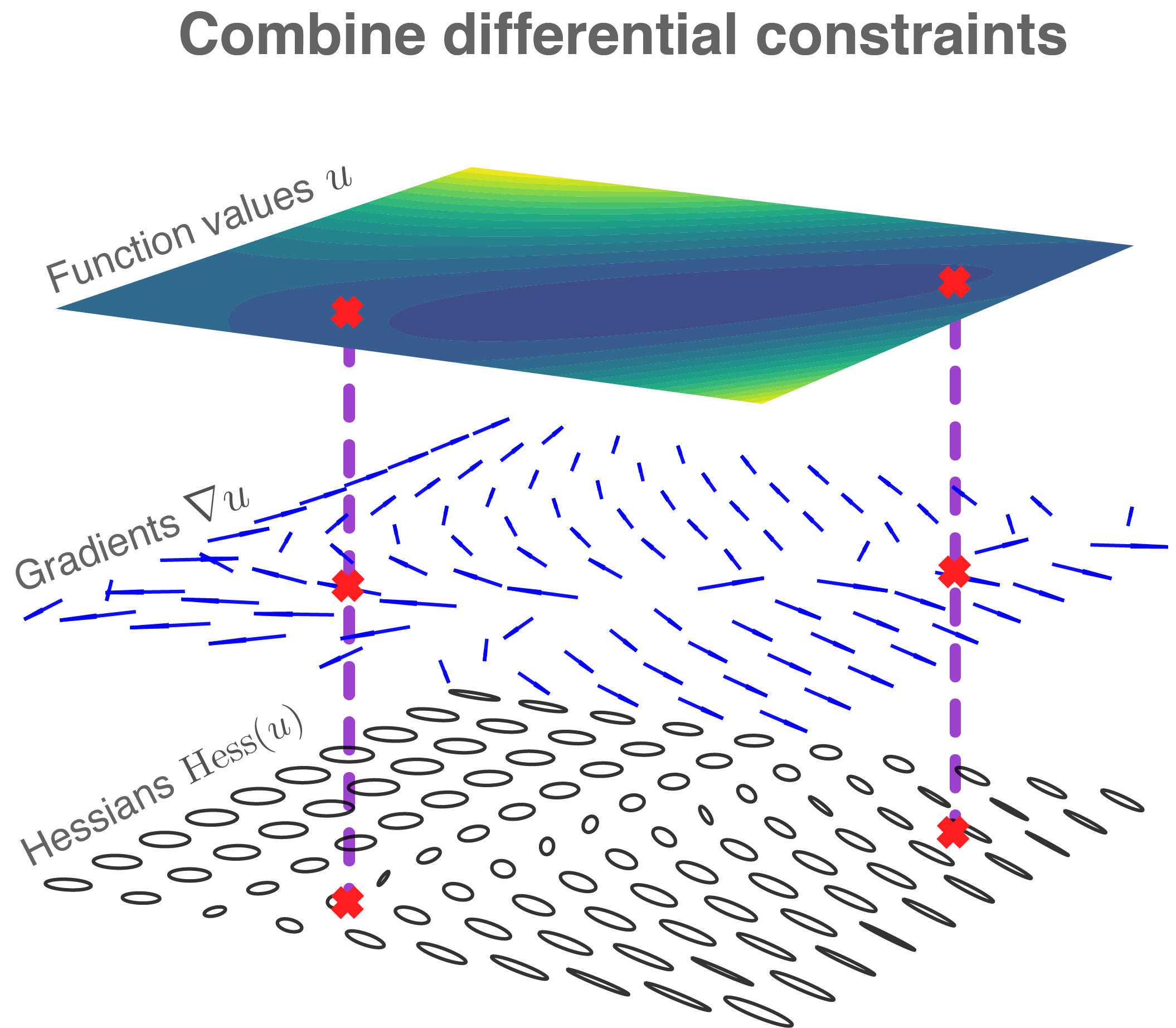}

\end{tocentry}

\begin{abstract}
Reconstructing force fields (FFs) from atomistic simulation data is a challenge since accurate data can be highly expensive. Here, machine learning (ML) models can help to be data economic as they can be successfully constrained using the underlying symmetry and conservation laws of physics. However, so far, every descriptor newly proposed for an ML model has required a cumbersome and mathematically tedious remodeling. We therefore propose using modern techniques from algorithmic differentiation within the ML modeling process -- effectively enabling the usage of novel descriptors or models fully automatically at an order of magnitude higher computational efficiency.
This paradigmatic approach enables not only a versatile usage of novel representations and the efficient computation of larger systems -- all of high value to the FF community -- but also the simple inclusion of further physical knowledge such as higher-order information (e.g.~Hessians, more complex partial differential equations constraints etc.), even beyond the presented FF domain.
\end{abstract}

Most physical quantities are represented by differential equations (DEs)
\begin{equation}
\mathcal{L} u(\mathbf{x}) = \mathbf{f}(\mathbf{x}),
\label{DE}\end{equation}
where $\mathcal{L} = \sum_{|j| \leq n} a_j \mathcal{D}^{j}$ is a finite linear combination of differential operators $\mathcal{D}^j$ of order $n$.

Examples of Eq.~(\ref{DE}) are found in electrodynamics, fluid dynamics, quantum mechanics etc.~always imposing  strong constraints on the form of a physical solution $\mathbf{f}(\mathbf{x})$. For several years,   machine learning (ML) has started to broadly contribute to physical modeling across its disciplines, for instance in particle physics \cite{Baldi2014}, atomistic simulations and force fields (e.g.~\cite{rupp2012fast,unke2020,von2020exploring,noe2020machine,behler2011atom}), fluid dynamics (e.g.~\cite{brunton2020machine,kochkov2021machine}) or lattice gauge theories~\cite{albergo2019flowbased,kanwar2020,nicoli2021estimation,luo2021gauge}. Note, however, that the standard procedure of ML models has been so far to learn from empirical data only. 
Recently, ML modeling has also started to incorporate physical constraints, be it as regularization terms (e.g.~\cite{karniadakis2021physics,pun2019physically}) or by explicitly taking into account conservation laws, DEs such as Eq.~(\ref{DE}), symmetries (e.g.~\cite{chmiela2017,chmiela2018, unke2020}), various operator response properties~\cite{christensen2019operators, unke2021se} or empirical correction terms, e.g. for long-ranged electrostatics~\cite{yao2018tensormol, unke2019physnet, grisafi2019incorporating, unke2021spookynet}. Notably, the direct usage of DEs allows the method to greatly simplify modeling and to be highly data efficient {\em since the known laws of physics do not need to be learned from empirical data anymore} (see e.g.~\cite{chmiela2017,chmiela2018, unke2020}). 

Let us study the popular example of Eq.~(\ref{DE}) force fields (FFs), namely the negative gradient $\mathcal{L} = -\nabla_{\mathbf{x}}$, where $u(\mathbf{x})$ is a scalar energy potential mapped to a vector field of forces $\mathbf{f}(\mathbf{x})$. The unknown force field (FF) is sampled at several locations $\left\{\mathbf{x}_{i}\right\}_{i=1}^{M}$ and $\mathbf{f}(\mathbf{x}_{i})$ are obtained from \emph{ab initio} atomistic reference data~\cite{bartok2015gaussianforce,chmiela2017,noe2020machine}. The constraint 
\begin{equation}
    \mathcal{L} = -\nabla_{\mathbf{x}}
\label{eq:force}
\end{equation}
 which is equivalent to both energy conservation and curl-free vector fields,
enables the estimation of a faithful representation of true energy-conserving FFs as opposed to the estimation of general (unphysical) vector fields~\cite{bartok2010gaussian, chmiela2017, christensen2020fchl19}. 

This approach unfolds its usefulness well beyond the FF domain for higher order differential operators and compositions thereof~\cite{batzVariationalEstimationDrift2016, sriperumbudurDensityEstimationInfinite, zhouNonparametricScoreEstimators2020,eriksson2018scaling,roosHighDimensionalGaussianProcess2021}, which characterize complex properties across all physics domains, e.g.  Hessians or Laplacian operators~\cite{sarkka2011linear, jidling2017linearly, finzi2020probabilistic}. Prior studies~\cite{learyDerivativeBasedSurrogate2004, osborneGaussianProcessesGlobal, solakDerivativeObservationsGaussian, giesl2018kernel, chmiela2017, chmiela2018, graepel2003SolvingNoisyLinear, langehegermann2021LinearlyConstrainedGaussian, raissi2017MachineLearningLinear, thomas2018tensor, mardt2018vampnets, greydanus2019hamiltonian, christensen2019operators, cranmer2020lagrangian, lutter2019deep,huang2022madgp} have demonstrated the effectiveness of explicit DE constraints, but their adoption still remains low due to an inefficiency that hampered model development and training so far: (1) Differential operator transforms inflate model complexity, making their full algebraic derivation mathematically tedious and error prone.
(2) Training and evaluation of a model using DE constraints is often associated with significant time and memory costs. While there exist prior works with focus on efficiency improvement, they rely on manual derivations and are limited to first-order gradients and primitive kernels without physical descriptors~\cite{eriksson2018scaling,roosHighDimensionalGaussianProcess2021,zhouNonparametricScoreEstimators2020}. On the other hand, recent existing work that focuses on automating the derivation of (first-order) constraints is not able to avoid a costly dense instantiation of the full model~\cite{huang2022madgp}.

\begin{figure*}
    \centering
    \includegraphics[width=\textwidth]{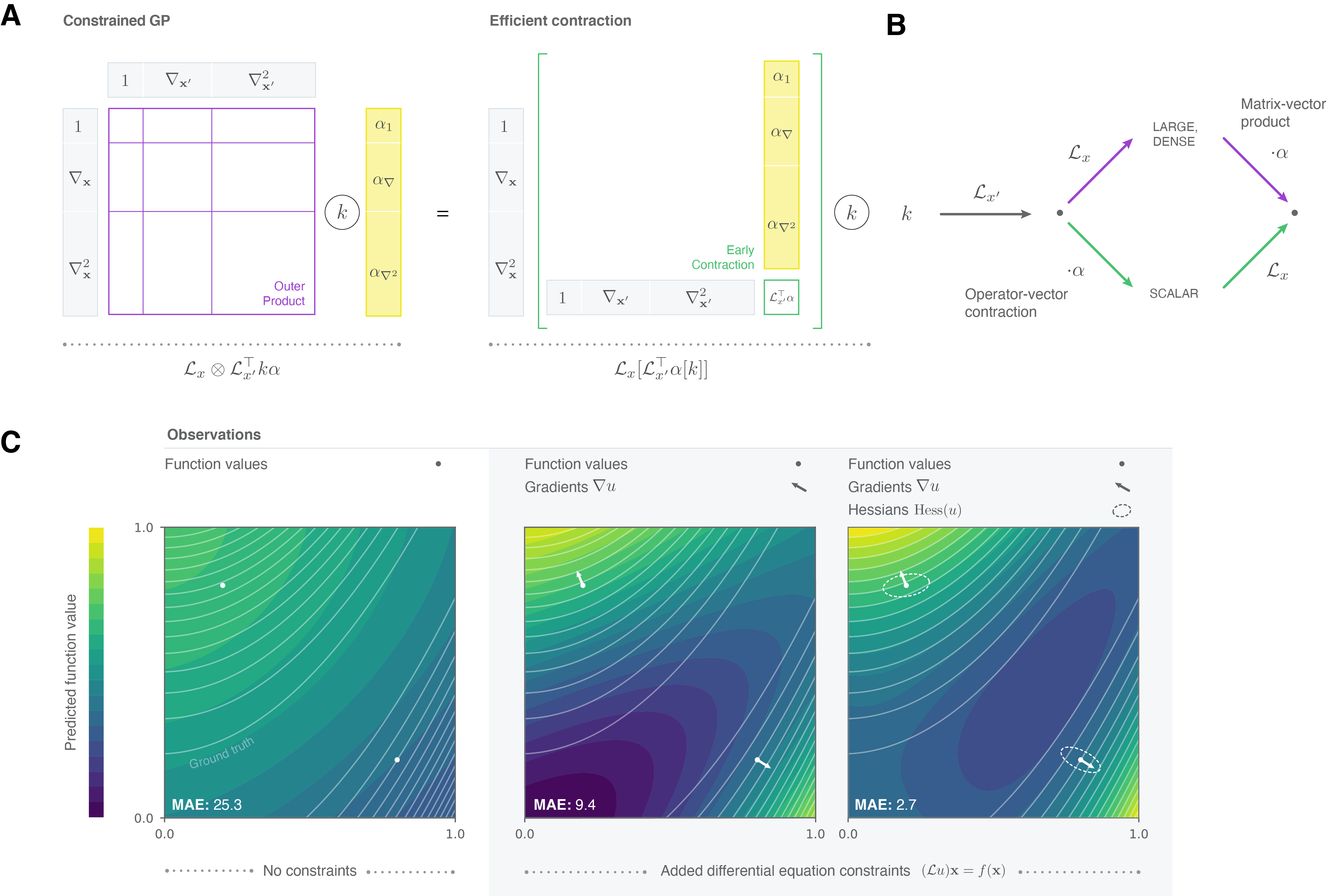}
    \caption{(A and B) More efficient kernel-vector products. Kernel functions subject to linear differential operator constraints have a tensor-product structure (purple), which can be leveraged by managing the contraction order when computing kernel-vector products. 
    Gaussian processes can benefit enormously from this improvement, since the full instantiation of the kernel matrix can be avoided during inference and training. (C) The effectiveness of differential equation constraints. A Gaussian process (RBF kernel, length scale $\sigma = 1$) is used to reconstruct the two-dimensional Rosenbrock function $u(x_1,x_2)=(1-x_1)^2+100(x_2-x_1^2)^2$ (white contour lines) from just two training points. Function value samples alone are insufficient to recover any surface features (left), but gradient constraints already enable the approximate recovery of extrema, despite not having been directly sampled (middle). Hessian constraints further refine the local curvature of the reconstruction (right).}
    \label{fig:overview}
\end{figure*}

In this work, we address both issues simultaneously using methods from algorithmic differentiation (AD)~(e.g. refs.~\cite{bryson1962steepest,griewank2008EvaluatingDerivatives,goodrich2021designing,baydin2018automatic}). While AD is now routinely used to automate the computation of derivatives, another benefit -- which turns out to be essential for this work -- is that it can simplify models that have a DE structure. As we will see, having a DE structure allows the collapse of certain portions of the computational graph {\em early}  during evaluation, by preaccumulating derivatives according to the chain rule. In summary, there are three key aspects that make this work useful for a broad set of applications in physical modeling:
\begin{itemize}
    \item It enables the systematic construction of empirical models subject to complex differential constraints.
    \item It is easy to use and versatile, even for elaborate neural network descriptors.
    \item It affords orders of magnitude gains in computational speed.
\end{itemize}
We demonstrate this for a number of popular FF models~\cite{unke2020} based on Gaussian process (GP) estimators under linear operator transformations. Such models can then be decomposed into tensor products of operators due to unique properties of the kernel function (see Fig.~\ref{fig:overview}).

As we will show, AD can yield efficient and automatic construction of GPs. 
Previously, a full instantiation of the DE constrained models was necessary, which due to the laborious manual derivations required,  warranted a separate publication for each new constrained model (e.g.~\cite{chmiela2017, chmiela2018, christensen2019operators, christensen2020fchl19, klus2021symmetric}). Using AD, we show that GPs can be created \emph{ad hoc} and \emph{automatically} (avoiding the above discussed manual derivation steps) for arbitrary DE constraints, as we demonstrate by reimplementing a broad selection of popular classic as well as new ML-FFs -- all  can now be  studied within our novel  framework.

This newfound efficiency also enables us to easily recombine promising concepts from existing models into new, even more powerful physics models.

Specifically, AD enables us to break down differential expressions into sets of linear primitive operations, which are evaluated one-by-one to avoid a full instantiation of all intermediates of $\mathcal{L}u(\mathbf{x})$, whenever the response is needed only for a specific transformation of $\mathbf{x}$. Due to the significant internal structure of derivatives, the individual terms that comprise $\mathcal{L}u$ are often low-dimensional and can thus be evaluated more efficiently by optimally managing the contraction order without having to undergo an unnecessary inflation.

Following this general idea, it turns out that most operator constraints can be applied with a surprisingly low overhead, often without even increasing the asymptotic computational complexity class of the original model. This fact is rather unexpected as it is unintuitive, since a constrained model often involves significantly larger terms, which are expensive if evaluated naively.
AD draws its efficiency from just two fundamental operations~\cite{griewank2008EvaluatingDerivatives}: Given a differentiable function $u:\R^N\to\R^L$ with total evaluation time cost $C_u$, AD guarantees that for $\mathbf x\in\R^N$, $\mathbf v\in\R^N$ and $\mathbf w\in\R^L$

\begin{enumerate}
    \item Jacobian-vector product $\mathbf{J}_{u}(\mathbf x)\mathbf v$ has  cost 
$\bigO(C_u)$, 
    \item Jacobian-transpose-vector product $\mathbf{J}_{u}^\top(\mathbf x) \mathbf w$ has  cost  $\bigO(C_u)$.
\end{enumerate}
Since any linear differential operator can be composed using these two simple rules, thus often at similar low cost (see Fig.~\ref{fig:overview} and Tables~\ref{tab:ad_contractions} and \ref{tab:ADoverview}), e.g. Hessian-vector-products in $\bigO(C_u)$ instead of quadratic complexity. These improvements are possible, because the full Jacobian never has to be calculated or stored.

Such inexpensive Jacobian products can be leveraged in any differentiable model, but they are particularly useful when applying DE constraints to GPs (for FFs) as we will now discuss 
\begin{equation}
u(\mathbf{x}) \sim \mathcal{G} \mathcal{P}\left({\mu}(\mathbf{x}), k\left(\mathbf{x}, \mathbf{x}^{\prime}\right)\right)
\end{equation}
is fully defined by a mean ${\mu}: \R^{n} \rightarrow \R$ and a covariance function $k(\cdot, \cdot): \R^{n} \times \R^{n} \rightarrow \R$, and is readily learned from data \cite{mackay1998introduction,rasmussen2001occam, rasmussen2004gaussian,rupp2012fast,bartok2010gaussian}. For instance many FFs, with $n=3N$ atomic degrees of freedom,
are modeled by GPs from \emph{ab initio} reference calculations (cf.~\cite{unke2020, keith2021combining,von2020exploring,noe2020machine}).

A natural benefit of GPs for DE-based regression is that they are closed under linear operators, leading to the form
\begin{align}
    \mathbf{f}(\mathbf{x}) \sim
    \mathcal{GP}\left(\mathcal{L}{\mu}(\mathbf{x}),\mathcal{L}^{}_{\mathbf{x}}\otimes\mathcal{L}^\top_{\mathbf{x}'} k\left(\mathbf{x}, \mathbf{x}^{\prime}\right)\right)
\end{align}
where subscripts indicate the argument of the kernel for each operator to act on.

As a concrete example, consider observing function values, gradients, and Hessians, i.e. $\mathcal{L} = (1, \nabla, \nabla^2)$ within the same model (cf. Fig.~\ref{fig:overview}). The corresponding kernel will take the form

\begin{align}
\mathcal{L}_{\mathbf{x}}^{}\otimes\mathcal{L}^\top_{\mathbf{x}'} k
=
\left[
\begin{array}{c|c|c}
k & \nabla_{\mathbf{x}'} k & \nabla^2_{\mathbf{x}'} k\\\hline
\nabla_{\mathbf{x}}k & \nabla_{\mathbf{x}}^{}\otimes\nabla_{\mathbf{x}'} k & 
\nabla_{\mathbf{x}}^{}\otimes\nabla^2_{\mathbf{x}'} k\\\hline
\nabla^2_{\mathbf{x}}k & \nabla^2_{\mathbf{x}}\otimes\nabla_{\mathbf{x}'}k & 
\nabla^2_{\mathbf{x}}\otimes\nabla^2_{\mathbf{x}'}k\\
\end{array}\right],\label{eq:gradhesskernel}
\end{align}
where each differential constraint appears as a cross-covariance combination with all other terms (e.g. the Hessian-Hessian covariance part being a fourth order derivative). We leverage AD to construct this exceedingly complex matrix without the need to instantiate the corresponding analytical expressions. Rather, each term is constructed efficiently by contracting local Jacobian-vector products on-the-fly (cf. Fig.~\ref{fig:overview}).

Likewise, for third-order operators, the kernel uses a sixth-order derivative, and so on. At this complexity growth rate, a manual algebraic derivation becomes increasingly hopeless.

But with the help of AD, a mere definition of $k (\mathbf x, \mathbf x')$ is sufficient, avoiding taking manual derivatives. Moreover computationally the action of the operator avoids an intermediate generation of the full matrix expression which gives rise to high computational efficiency.

We would like to discuss  further the efficiency aspect of our AD framework, which also goes beyond standard AD usage. 
While the standard computation of the GP equations, for example, for the predictive mean are 
${\mathbf{f}(\mathbf{x}) = \sum_i \mathcal{L}^{}_{\mathbf{x}}\otimes\mathcal{L}^\top_{\mathbf{x}_i} k\left(\mathbf{x}, \mathbf{x}_i\right) \bm{\alpha}_i}$, 
where coefficients $\bm{\alpha}_i$ are obtained by solving a linear system (see Eqs.~\ref{eq:S2},~\ref{eq:S3} and {refs.}~\cite{rasmussen2004gaussian,rupp2012fast}),
we note that this computation is usually dense and therefore slow. 

Due to linearity, we can alternatively compute the same at roughly an order more efficiently (see Fig.~\ref{fig:timings} and Table~\ref{tab:ad_contractions}) by
\begin{align}
\mathbf{f}(\mathbf{x}) = \sum_i \mathcal{L}^{}_{\mathbf{x}}\left[\mathcal{L}^\top_{\mathbf{x}_i} \bm{\alpha}_i k\left(\mathbf{x}, \mathbf{x}_i\right)\right]
\end{align}
as the operator tensor product is resolved at first into a scalar operator
$\mathcal{L}^\top_{\mathbf{x}_i}\bm{\alpha}_i$
which can take full advantage of the automatic contraction rules inherent to AD. For an intuitive explanation see Fig.~\ref{fig:overview} and  for a more detailed derivation see the Supporting Information.

We will now apply the AD framework discussed above to the atomistic simulations domain. For this we proceed with the construction of  potential energy surfaces (PESs) for small molecules from the well-established MD17 benchmark data set~\cite{chmiela2017}, using various GP-based ML-FFs that we recreate within our novel framework. First, we consider the Symmetric Gradient-domain Machine Learning (sGDML)~\cite{chmiela2018, chmiela2022} model, which uses derivative constraints (Eq.~(\ref{eq:force})) to simultaneously reconstruct a conservative molecular FFs and their corresponding PESs. This model employs a twice differentiable kernel function from the parametric Mat\'{e}rn family~\cite{matern1986spatial, Gradshteyn2007, Gneiting2010}, which is symmetrized to be invariant with respect to the relevant rigid space group, as well as dynamic nonrigid symmetries of the system at hand. It is then combined with a descriptor that enumerates all unique pairwise inverse distances between atoms. This composition of functions -- if implemented in the standard manner \cite{chmiela2019} -- leads to an increased cost since all atomic degrees of freedom of the molecule enter the model as separate constraints. 
Our AD framework resolves this complexity and avoids unnecessary instantiation of operator tensor products, yielding an order $3N$ improvement (where $N$ is the number of atoms) in the case of gradient operators. 

Furthermore, we have reimplemented other FF GP-based models, including FCHL19~\cite{FCHL2018, christensen2020fchl19}. %
 Within our AD framework, we could do so by specifying the respective kernel functions without needing to manually implement derivatives. 
 These models are compared to (s)GDML models trained on the exact same data set splits. To verify that our own implementations are correct, we have compared our test errors with the respective original publications. 

Analyzing running times of the highly optimized FCHL19 \cite{christensen2020fchl19} reference implementation with our own AD based one, we can already see the high intrinsic optimization abilities and computational advantages for our AD framework, which is able to generate predictions up to two orders of magnitude faster, while using only a single GPU instead of a CPU cluster node (cf. Table~II in~\cite{christensen2020fchl19} and Table~\ref{tab:timings}).
We note, however, as a word of caution to this comparison that different compute architectures and programming languages are being used. 

Table~\ref{tab:timings} contains a running time comparison of all considered models within our AD framework for the largest (aspirin) and smallest (ethanol) molecule in the MD17 data set, both with and without relying on early contraction of intermediates. Since AD alleviates the need for laborious manual derivations, we can easily replicate further  ML-FFs within the same code to arrive at  fair running time comparisons. This type of analysis could not be done accurately before.

The ultimate advantage of our AD framework is the freedom to recombine promising concepts from existing approaches effortlessly on a broad scale, in order to discover even more powerful task-specific models. So far, there exists no single universally best ML-FF; the field is rather characterized by specialized solutions. We demonstrate this new modeling flexibility by creating several better variants of the GP-based models mentioned above. \emph{sGDML[RBF]} is a sGDML variant that uses the RBF kernel instead of the Mat\'{e}rn kernel. \emph{global-FCHL19} is a global variant of FCHL19, which parametrizes each atom interaction individually at the cost of permutational invariance. And finally, \emph{sFCHL19} is a symmetrized global variant using the symmetries recovered by the sGDML model. Table~\ref{tab:remix_models} compares these variants to the respective original models and demonstrates that highly significant advances in prediction accuracy become possible for all MD17 data sets, simply, by systematically combining existing ideas. 
Even linearly combining kernels shows consistent improvements (see Table~\ref{tab:remix_models2}). Furthermore, AD allows an effortless gradient-based hyperparameter
optimization, as we demonstrate in section~\ref{ss:hyper_opt} in the Supporting Information.

\begin{table*}
    \centering
    \begin{adjustbox}{max width=\textwidth}
    \begin{tabular}{l|cccccccc}
    \toprule
                    \textbf{Model} & Aspirin  & Benzene  & Ethanol  & Malonaldehyde    & Naphthalene   & Salicylic acid     & Toluene  & Uracil\\
    \midrule
    sGDML           & 0.702 & 0.163 & 0.341 & 0.410 & 0.115 & 0.286 & 0.145 & 0.242\\
    FCHL19GPR       & 0.628 & 0.179 & 0.180 & 0.302 & 0.185 & 0.277 & 0.246 & 0.147 \\
    \midrule 
    sGDML[RBF]       & 0.506 & \textbf{0.147} & 0.186 & \textbf{0.237} & \textbf{0.068} & \textbf{0.167} & \textbf{0.096} & 0.140 \\
    global-FCHL19   & 0.748 & 0.235 & 0.319 & 0.429 & 0.292 &~ 0.216 & 0.370 & \textbf{0.120} \\
    sFCHL19 & \textbf{0.430} & 0.158 & \textbf{0.160} & 0.274 & 0.094 & 0.172 & 0.111 & 0.130 \\
    \bottomrule
    \end{tabular}
    \end{adjustbox}
    \caption{\textbf{Test errors (MAE) for force learning on MD17 with 1000 reference points using various new GP-FF variants.} All errors in kcal mol$^{-1}$ Å$^{-1}$. Best results are in bold. The automation of AD allowed us to recombine existing ideas from different ML-based FFs on a broad scale and find better performing model variants for all MD17 data sets. The best result for each data set is marked in bold.}
    \label{tab:remix_models}
\end{table*}
We can even go a step further and use the representations learned by modern deep architectures~\cite{SchNet2018,schutt2021equivariant,wilson2016deep} as pretrained descriptors $\mathbf{D}(\mathbf{x})$ within GDML-type models. Each descriptor then yields a new composite kernel
\begin{equation}
\tilde{k}(\mathbf{x}, \mathbf{x}') = k\left(\mathbf{D}(\mathbf{x}), \mathbf{D}(\mathbf{x}')\right).
\end{equation}
Our numerical results show that this construction interestingly yields models that  are more accurate than their (pretrained) ingredients. Table~\ref{tab:remix_nn} summarizes the force prediction performances, showing that significant improvements are possible following this simple strategy. {We reiterate that this advance was enabled by the unprecedented flexibility to remix different models enabled by our framework.}
 
In order to demonstrate that the AD framework can also be applied beyond gradient observations (constraint from Eq.~(\ref{eq:force})), we will now illustrate the effect of employing more complex higher-order DE constraints, namely, Hessians for the learning model. 

Using the Hessian-kernel in Eq.~(\ref{eq:gradhesskernel}), we reconstruct a toy surface from just two gradient/Hessian observations (see Fig.~\ref{fig:overview}). Notably, this example shows that constraints with higher derivatives are more informative and aid regression models significantly. In our example, the reconstruction error improves by one order of magnitude as a result of including second-order measurements. With that, even  a small sample size becomes sufficient to identify the correct model. For further illustrative examples using other DE constraints and ML-based DE solving, see Table~\ref{tab:ad_contractions}. 

\begin{table*}
    \centering
    \begin{adjustbox}{max width=\textwidth}
    \begin{tabular}{ll|cccccccc}
        \toprule
        \textbf{Model} & \textbf{Descriptor}     & Aspirin        & Ethanol        & Malonaldehyde  & Naphthalene     & Salicylic acid & Toluene        & Uracil         \\
        \midrule
        SchNet         & -                       & \textbf{0.824} & 0.225          & 0.428          & 0.342          & 0.490          & 0.325          & 0.307          \\
        sGDML[RBF]     & SchNet                  & 0.930          & \textbf{0.176} & \textbf{0.359} & \textbf{0.294} & \textbf{0.459} & \textbf{0.281} & \textbf{0.248} \\
        \midrule
        PaiNN          & -                       & \textbf{0.389} & 0.220          & 0.336          & 0.103          & \textbf{0.232} & 0.122          & \textbf{0.176} \\
        sGDML[RBF]     & PaiNN (scalar features) & 0.422          & \textbf{0.186} & \textbf{0.321} & \textbf{0.102} & 0.241          & \textbf{0.118} & 0.182          \\
        \bottomrule
    \end{tabular}
    \end{adjustbox}
    \caption{\textbf{Using molecular representations generated by various deep neural network architectures as descriptor for GDML-type models.} All models have been trained on the same 1000 reference points for each molecule in the MD17 data set. All (force) test errors (MAE) are in kcal mol$^{-1}$ Å$^{-1}$. The best result for each data set is marked in bold.}
    \label{tab:remix_nn}
\end{table*}

\begin{table*}
\centering
    \begin{adjustbox}{max width=\textwidth}
    \begin{tabular}{l|ccc|ccc}
    \toprule
                    \textbf{Model} & \multicolumn{3}{c}{\textbf{Ethanol ($N=9$)}} & \multicolumn{3}{c}{\textbf{Aspirin ($N=21$)}} \\
                    & dense & contraction & speedup & dense & contraction & speedup \\
    \midrule
    sGDML           & \phantom{0}0.0780 & 0.0018 & $\times$\textbf{43.3}& \phantom{00}0.2832 & \phantom{0}0.0087 & $\times$\textbf{32.5}\\
    FCHL19GPR       & 57.9439 & 0.9815 & $\times$\textbf{59.0} & 414.0216 & 10.1438 & $\times$\textbf{40.8}\\
    \midrule
    sGDML[RBF]      & \phantom{0}0.0800 & 0.0015 & $\times$\textbf{53.3}& \phantom{00}0.2811 & \phantom{0}0.0074  & $\times$\textbf{37.9}\\
    global-FCHL19   & 57.5194 & 0.9653  & $\times$\textbf{59.5} & 458.5465 & 10.0728 & $\times$\textbf{45.5}\\
    sFCHL19         & 60.3951 & 1.0704  & $\times$\textbf{56.4} & 419.3778 & 10.3336  & $\times$\textbf{40.5}\\
    \bottomrule
    \end{tabular}
    \caption{Benchmarked force prediction times (s) for different kernels. Each model was trained using 1000 points, and evaluated for a batch of 10 points. All timings are averaged over 10 runs (excluding an initial run for just-in-time compilation). Our approach (contraction) yields consistent speedups by a up to two orders of magnitude over the direct (dense) implementation of the constrained models. All measurements are done on a single Nvidia Titan RTX 24 GB GPU.}
    \label{tab:timings}
    \end{adjustbox}
\end{table*}

In summary, an overwhelming number of physical phenomena are governed by linear DEs that can be used as highly effective physical knowledge in data-driven estimation problems. By excluding physically infeasible solutions, such constraints play a crucial role in obtaining data-efficient and robust models. We have used methods from AD to address two key challenges that have so far hindered widespread adoption of this approach: (1)  the full algebraic instantiation of the constrained model is expensive (or even unfeasible) to train and evaluate  and (2) the application of DEs to GPs by hand is tedious and error prone for a modeler. Due to these obstacles, the construction of DE-constrained models did not leave much room for explorative and swift model creation in the past. To change that, we have contributed how our AD framework can be used to automate the model construction process and how regularities in the differential structure can be leveraged to gain efficiency in practice. {This general framework has enabled us to construct and integrate various differential operator constraints that constitute the basic building blocks of most physical laws into ML-FFs.} 
Following the same principle, more complex operators can be constructed and turned into constrained ML models; we used mainly GPs to show this point. For further examples, see the Supporting Information.

Finally, we have demonstrated in a series of numerical experiments, how our framework can be used to readily replicate some state-of-the-art DE-constrained GP-based FFs  by simply changing a few lines of code without the need to go through tedious manual derivations. Going one step further, we were able to demonstrate how competitive new kernelized variants of existing deep learning-based FFs can be developed and combined.

\begin{acknowledgement}

This work was supported in part by the German Ministry for Education and Research (BMBF)
under Grants 01IS14013A-E, 01GQ1115, 01GQ0850, 01IS18056A, 01IS18025A and 01IS18037A, 
 by the
Information \& Communications Technology Planning \& Evaluation (IITP) grant funded by the Korea government (No. 2017-0-001779,  Artificial Intelligence Graduate School Program, Korea University). %
We thank OT Unke for very helpful comments on the manuscript. Furthermore, we thank IPAM for warm hospitality and inspiration while finishing the manuscript. Correspondence to KRM and SC. 

\end{acknowledgement}

\begin{suppinfo}

Further details about computational settings and benchmarks, toy examples, derivations of computational complexity considerations, Tables SI-SIV and Figures S1 and S2, as well as source code to all experiments.

\end{suppinfo}

\bibliography{ms}

\clearpage
\pagebreak

\appendix

\setcounter{equation}{0}
\setcounter{figure}{0}
\setcounter{table}{0}
\setcounter{page}{1}
\makeatletter
\renewcommand{\thepage}{S\arabic{page}}
\renewcommand{\theequation}{S\arabic{equation}}
\renewcommand{\thefigure}{S\arabic{figure}}
\renewcommand{\thetable}{S\Roman{table}}
\twocolumn[
\begin{@twocolumnfalse}
\begin{center}
\textbf{\large Supplementary Information for ``Algorithmic Differentiation for Automated Modeling of Machine Learned Force Fields''}\newline\newline
Niklas Frederik Schmitz,$^\dag$ Klaus-Robert M\"uller,$^{*,\dag,\ddagger,\P,\S,\parallel}$ and Stefan Chmiela$^{*,\dag,\ddagger}$\newline\newline
$\dag$\emph{Machine Learning Group, Technische Universität Berlin, 10587 Berlin, Germany}\\
$\ddagger$\emph{BIFOLD - Berlin Institute for the Foundations of Learning and Data, 10587 Berlin, Germany}\\
$\P$\emph{Department of Artificial Intelligence, Korea University, Seongbuk-gu, Seoul 02841, Korea}\\
$\S$\emph{Max Planck Institute for Informatics, Stuhlsatzenhausweg, 66123 Saarbrücken, Germany}\\
$\parallel$\emph{Google Research, Brain Team, 10117 Berlin, Germany}\newline\newline
E-mail: klaus-robert.mueller@tu-berlin.de; stefan@chmiela.com
\end{center}
\end{@twocolumnfalse}
]

\section{\label{s:ad_kernels}AD boosts the efficiency of constrained kernel evaluations for GPs}
GPs are closed under linear transformations and can therefore naturally represent {functions under} linear operators on %
continuous domains as
\begin{equation}
\mathcal{L} u \sim \mathcal{G} \mathcal{P}\left( \mathcal{L} \mu(\mathbf{x}), \mathcal{L}^{}_{\mathbf{x}}\otimes\mathcal{L}^\top_{\mathbf{x}'}
k\left(\mathbf{x}, \mathbf{x}^{\prime}\right)\right),
\end{equation}
where subscripts indicate the argument of the
kernel for each operator to act on. Our focus lies on the definition of the constrained kernel function $\mathbf{K}_{\mathcal{L}, \mathcal{L}}(\mathbf x, \mathbf{x}^{\prime}) = \mathcal{L}^{}_{\mathbf{x}}\otimes\mathcal{L}^\top_{\mathbf{x}'}
k\left(\mathbf{x}, \mathbf{x}^{\prime}\right) = \left[\mathcal{L}_a\mathcal{L}_b k\left(\mathbf{x}, \mathbf{x}^{\prime}\right)\right]_{ab}$, which is commonly implemented as a dense matrix by 
naively evaluating the aforementioned expression without further considerations. The problem with this direct approach is that the model grows with the number of partial derivative constraints, thus inflating training and inference cost dramatically. To avoid this, we retain the tensor product view, which makes the expression amenable to optimization during evaluation. Any differential operator within the kernel is decomposed into Jacobian-vector products and contracted efficiently. Table~\ref{tab:ad_contractions} outlines good decompositions for the most important elementary differential operators applied to kernels and shows the computational saving that can be achieved (see Table~\ref{tab:ADoverview} for further details). AD also exposes low-rank structures and sparsity patterns in the expression to further boost the contraction speed~\cite{griewank2008EvaluatingDerivatives} (see Fig.~\ref{fig:overview}).
This combination of mechanisms is exceedingly effective in GPs, which only ever evaluate the kernel function in {\em one} direction of parameter space ${\bm{\alpha}}_i$ at a time, as
\begin{equation}
\mathcal{L} u(\mathbf x) = \sum_i \mathbf{K}_{\mathcal{L}, \mathcal{L}}(\mathbf x, \mathbf x_i) \mathbf {\bm{\alpha}}_{i}.
\label{eq:S2}
\end{equation}
{The coefficients $\bm{\alpha}$ are obtained from solving the linear system $(\mathbf{K}_{\mathcal{L}, \mathcal{L}}+\lambda \mathbb{I})\bm{\alpha} = \mathbf{F}$, 
where $\mathbf{K}_{\mathcal{L}, \mathcal{L}}=[\mathbf{K}_{\mathcal{L}, \mathcal{L}}(\mathbf{x}_i, \mathbf{x}_j)]_{ij}^m$ is the kernel matrix computed from a training set $\{(\mathbf{x}_i, \mathbf{f}(\mathbf{x}_i))\}_{i=1}^m$, $\mathbf{F}=[\mathbf{f}(\mathbf{x}_i)]_{i}$ and $\lambda > 0$ a regularization parameter.} %
Furthermore, fast kernel-vector products are not only useful for model evaluation, but also training can be expressed as a repeated evaluation of kernel-vector products by iterating
\begin{equation}
\mathbf{\bm{\alpha}}^{t}=\bm{\alpha}^{t-1}-\gamma\left[(\mathbf{K}_{\mathcal{L}, \mathcal{L}}+\lambda \mathbb{I}) \mathbf{\bm{\alpha}}^{t-1}-\mathbf{f}(\mathbf{x})\right],
\label{eq:S3}
\end{equation}
where $\gamma$ is a learning rate. This view naturally carries over to other more sophisticated iterative linear solvers such as conjugate gradient methods, which only rely on vector products~\cite{chmiela2022}.

\begin{table*}[b]
\centering
\begin{tabular}{ll|lr|lr}
\toprule
$\mathcal{L}^{}_{\mathbf{x}}$ & $\mathcal{L}_{\mathbf{x}'}$ 
& $\left[\mathcal{L}_{\mathbf{x}}^{}\otimes \mathcal{L}_{\mathbf{x}'}^\top k(\mathbf{x},\mathbf{x}')\right]$ & cost
& $\mathcal{L}^{}_{\mathbf{x}}\left[\mathcal{L}^\top_{\mathbf{x}'}\bm{\alpha}\left[ k(\mathbf{x}, \mathbf{x}')\right]\right]$  & cost\\
\midrule
$\nabla$   & $\nabla$   & $\Jacobian_{\mathbf{x}}(\Grad_{\mathbf{x}'}(k))^\top$ & $NC_k$   & $\Grad_{\mathbf{x}}(\Grad_{\mathbf{x}'}(k)\bm{\alpha})$ & $C_k$\\
$\nabla$   & $\nabla^2$ & $\Jacobian_{\mathbf{x}}(\Hess_{\mathbf{x}'}(k))^\top$ & $N^2C_k$ & $\Grad_{\mathbf{x}}(\Hess_{\mathbf{x}'}(k) \bm{\alpha})$ & $NC_k$\\
$\nabla^2$ & $\nabla^2$ & $\Hess_{\mathbf{x}}(\Hess_{\mathbf{x}'}(k))^\top$     & $N^3C_k$ & $\Hess_{\mathbf{x}}(\Hess_{\mathbf{x}'}(k) \bm{\alpha})$ & $N^2C_k$\\
\bottomrule
\end{tabular}
\caption{Computational costs of common operator kernel instantiations compared to their contractions, assuming access to efficient reverse-mode AD. Here, $C_k$ denotes the cost of evaluation of the scalar base kernel $k(\mathbf{x},\mathbf{x'})$ for $\mathbf{x},\mathbf{x'}\in\mathbb{R}^N$. The coefficients $\bm{\alpha}$ always have the same size as the right operator $\mathcal{L}_{\mathbf{x}'}$. Note that the improvement in cost by our AD framework is a {\em factor} of $N$(!), where $N$ is the dimensionality of $\mathbf{x}$. }
\label{tab:ad_contractions}
\end{table*}

\section{Computational complexity}
\begin{table*}
    \centering
    \begin{tabular}{lrll}
    \toprule
            operation                  & time cost  & AD composition                & comment \\
    \midrule
    $\JVP_x(f, v)\in\R^m$              & $C_f$ & (primitive)                   &\\
    $\VJP_x(f, w)\in\R^n$              & $C_f$ & (primitive)                   &\\
    \midrule
    $\Jacobian_x(f)\in\R^{m\times n}$  & $\min\{n,m\}C_f$ & $[\JVP_x(f, e_i)\quad \textrm{for}\,i=1...n]$                    & or m VJPs\\
    $\Grad_x(f)\in\R^n$                & $C_f$ & $\VJP_x(f, 1)$                & for $m=1$\\
    $\Hess_x(f)\in\R^{n\times n}$      & $nC_f$ & $\Jacobian_x(\Grad_x(f))$       & for $m=1$\\
    $\Hess_x(f)\in\R^{m\times n\times n}$      & $mnC_f$ & $\Jacobian_x(\Jacobian_x(f))$       & \\
    $\Divergence_x(f)\in\R$            & $nC_f$ & $\trace(\Jacobian_x(f))$      & for $n=m$\\
    $\Laplacian_x(f)\in\R$             & $nC_f$ & $\Divergence_x(\Grad_x(f))$     &\\
    $\HVP_x(f,v)\in\R^n$                & $C_f$ & $\JVP_x(\Grad_x(f), v)$          &\\
    \bottomrule
    \end{tabular}
    \caption{Overview of selected differential operators and their respective known worst-case time complexities guaranteed by AD, assuming access to Forward mode (JVP) and Reverse mode (VJP) programming primitives. 
    {Here, HVP is the Hessian-vector product, and $e_i=(0,...,1,...,0)^\top$ denotes the $i-th$ standard basis vector in $\mathbb{R}^n$.} 
    We assume the primal function $f:\R^n\to\R^m$ to be implemented by a differentiable computer program with evaluation time cost $C_f$. We denote the input $x\in\R^n$ and auxiliary vectors $v\in\R^n, w\in\R^m$.}
    \label{tab:ADoverview}
\end{table*}

The efficient calculation of numerical values from differential operators $\mathcal L$ applied to functions $f$ at inputs $\mathbf{x}$ is a key challenge in algorithmic differentiation. From the building blocks of JVPs (Jacobian-vector products) and VJPs (Vector-Jacobian products) one can construct other common operators as exemplarily shown in Table~\ref{tab:ADoverview}. Importantly, the time cost is usually characterized in terms of the time cost of the underlying function, $C_f$, as well as input/output dimensions $n$ and $m$, as shown in the second column of Table~\ref{tab:ADoverview}. Understanding the explicit dependencies on $C_f,m,n$ is key to understanding the computational implications of AD application: Analyzing time complexity of nested  operators $\mathcal L_1\left[\mathcal L_2\left[.\right]\right]$, as well as characterizing the efficiency gain in our proposed kernel contraction.

We therefore analyze the time complexity in the kernel regime under linear operators, involving 
tensor products of operators such as $\mathcal{L}_{\mathbf{x}}^{}\otimes \mathcal{L}_{\mathbf{x}'}^\top k(\mathbf{x},\mathbf{x}')$.
For notation, we fix two linear operators ${\mathcal L_{\mathbf{x}}^{}: C^\infty(\mathcal X, \R)\to C^\infty(\mathcal X, \R^{d_1})}$ and ${\mathcal L_{\mathbf{x}'}^{}: C^\infty(\mathcal X, \R)\to C^\infty(\mathcal X,\R^{d_2})}$ of dimensionality $d_1$ and $d_2$, respectively.
The operator tensor product acting on the kernel can be evaluated in several ways, that have different efficiencies:

(A) The most basic strategy is to compute $d_1 \times d_2$ matrix entries independently, without sharing common intermediates. (B) A more refined approach is to take advantage of common sub-expressions when instantiating the full dense matrix (see Table~\ref{tab:ad_contractions}, column 1). (C) Our highly efficient AD approach is to focus on the whole expression in its tensor vector-product form, which allows to rearrange terms, e.g.~for an intermediate contraction of the (possibly high-dimensional) right operator $\mathcal{L}_{\mathbf{x}'}^\top$ into an operator $\mathcal{L}_{\mathbf{x}'}^\top\bm{\alpha}$ (see Table~\ref{tab:ad_contractions}, column 2). This yields a low dimensional result which strongly alleviates also all consecutive operator evaluations. 

Our approach is particularly effective whenever high-dimensional operations are involved in the construction of the operator. Examples of resulting drastic savings for operator kernel contraction are in Tab.~\ref{tab:ad_contractions}.

\section{Running times}
\begin{figure}
    \begin{adjustbox}{max width=\textwidth}
    \includegraphics[width=0.45\textwidth]{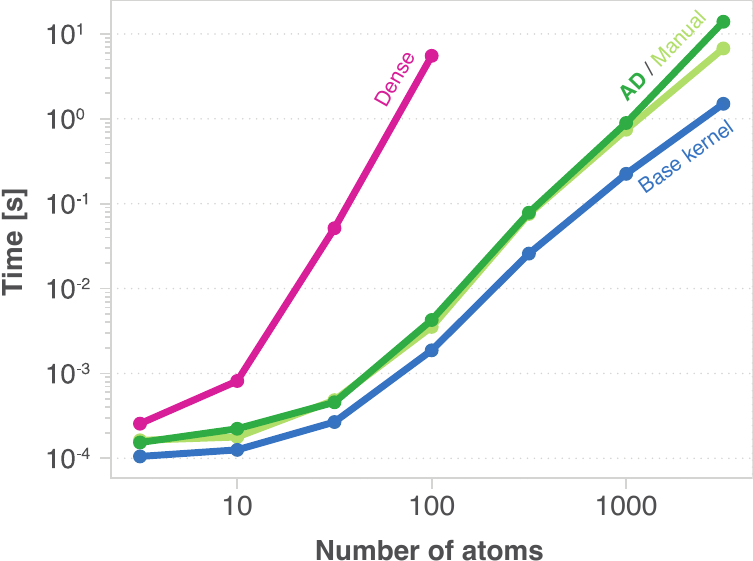}
    \end{adjustbox}
    \caption{\textbf{Efficient handling of differential equation constraints.} Empirical scaling of the prediction cost using different evaluation strategies for the GDML model trained on 1000 points, for systems with varying numbers of atoms. The baseline is given by the unconstrained scalar base kernel before applying any operator (blue). While the naive dense application of the operator scales poorly with increasing system size (pink), our fast AD implementation (green) always stays within only constant overhead to the scalar baseline. The running time of the AD implementation is on par with a manually derived optimal evaluation of the model (light green). We see improvements of two to four orders of magnitude for systems with 100 atoms.}
    \label{fig:timings}
\end{figure}

Timings for the GDML kernel under the gradient operator tensor product $\nabla^{}_{\mathbf{x}} \otimes \nabla^\top_{\mathbf{x}'}$ and respective speedups using our contraction approach 
are collected in Fig.~\ref{fig:timings}.
A comparison of force prediction timings of trained energy-conserving GPs for different base kernels is given in Table~\ref{tab:timings}.

\section{Gradient-based hyperparameter optimization}
\label{ss:hyper_opt}
AD can also be applied to jointly optimize any model hyperparameter using a gradient-based iterative scheme, while training the linear regression coefficients of the GP. Traditionally, hyperparameters are determined via computationally expensive grid searches, due to the difficulty to obtain gradients.

To demonstrate gradient-based hyperparameter optimization using AD, we generalize the molecular descriptor used in sGDML,
$$
D_{ij} = 1\: /\: ||R_i - R_j||,
$$
by introducing an exponent parameter $p$ on the pairwise atomic distances (default $p=1$),
$$
D_{ij} = 1\: /\: ||R_i - R_j||^p.
$$
We apply this extended descriptor in our best performing model variant (sGDML[RBF], see Table~\ref{tab:remix_models}), which also includes a variable length scale $\sigma$. This yields another model variant, sGDML[RBF,p], with two additional hyperparamters that need to be optimized.

In our numerical experiment, the regularization parameter $\lambda=10^{-10}$ is kept fixed. We take a small training set of 200 training points, which we partition into 80\%-20\% splits for the GP fit and validation force MAE evaluation, respectively. We then minimize the force MAE with respect to both hyperparameters with initializations $p=1$ and $\sigma=N$, where $N$ is the number of atoms, using the gradient-based Adam optimizer with step size $0.1$ for a fixed number of 200 steps. Finally, we re-fit the model on the full 200 training points with the found hyperparameters. For each molecule in MD17, the 200-step hyperparameter optimization loop takes less than 3 minutes in total on a single Nvidia A100 40GB GPU. The optimized parameters and results are shown in Table~\ref{tab:hypercoulomb}. The parametrically more complex sGDML[RBF,p] model variant outperforms the baseline sGDML[RBF] in all instances, except for the salicylic acid dataset, where we assume that the hyperparameter optimization converged to a local minimum.
\begin{table*}
    \centering
    \begin{adjustbox}{max width=\textwidth}
    \begin{tabular}{lcccccccc}
	\toprule
	                              & Aspirin        & Benzene        & Ethanol        & Malonaldehyde  & Naphthalene     & Salicylic acid & Toluene        & Uracil         \\
	\midrule
    $p$            &\phantom{0}1.361   & 0.695   & 0.102   & 0.238         & 0.861      &\phantom{0}1.559 & 0.564   & 0.754 \\
    $\sigma$       &29.798             & 2.744   & 0.329   & 0.765         & 2.364      &14.276           & 1.763   & 1.157 \\
    \midrule
    sGDML[RBF,p]      &\textbf{\phantom{0}1.310}   & \textbf{0.161}   & \textbf{0.450}   & \textbf{0.622}         & \textbf{0.137}      &\phantom{0}0.857 & \textbf{0.227}   & \textbf{0.429} \\
    \midrule
    sGDML[RBF] & \phantom{0}1.460 & 0.166 & 0.652 & 0.766 & 0.142 & \textbf{\phantom{0}0.699} & 0.259 & 0.440\\
	\bottomrule
    \end{tabular}
    \end{adjustbox}
    \caption{\textbf{Optimizing kernel hyperparameters of sGDML[RBF,p] by gradients.} Test errors (MAE) and found hyperparameters for force learning on MD17 with 200 reference points. All errors in kcal mol$^{-1}$ Å$^{-1}$. The best result for each dataset is marked in bold.} %
    \label{tab:hypercoulomb}
\end{table*}
\section{GPs using NNs as descriptors}
\label{ss:NN_descriptors}
Following the GDML approach, we construct energy-conserving GPs using $\mathcal{L} = \nabla$, which use
\begin{equation}
\nabla^{}_{\mathbf{x}}\otimes\nabla_{\mathbf{x}'}^\top\tilde{k}
=
\mathbf{J}^\top_{\mathbf{D}}
\left[\nabla^{}_{\mathbf{D}}\otimes\nabla^\top_{\mathbf{D}'}k\right]
\mathbf{J}^{}_{\mathbf{D}^{\prime}}
\end{equation}
as kernel functions. Here, $\mathbf{J}_{\mathbf{D}}$ are the Jacobians of the descriptor, according to the chain rule. Since the expression for $\mathbf{D}(\mathbf{x})$ is exceedingly complex in the case of NNs, it is highly impractical to derive these Jacobians by hand. Using AD however, these derivatives become straightforward to evaluate, which allows us to combine the best attributes of constrained GPs and deep architectures within one model~\cite{wilson2016deep}. For example, deep architectures excel at learning molecular representations, whereas GPs have a well-defined asymptotic behaviour and are easier to train due to the availability of closed-form solutions. We have considered the representations generated by SchNet~\cite{SchNet2018} 
and PaiNN~\cite{schutt2021equivariant} as descriptors, which we have combined with GPs that impose additional GDML-type energy-conservation and permutational symmetry constraints.

\section{Numerical experiments (toy examples)} 

\begin{figure*}
    \centering
    \includegraphics[width=0.7\textwidth]{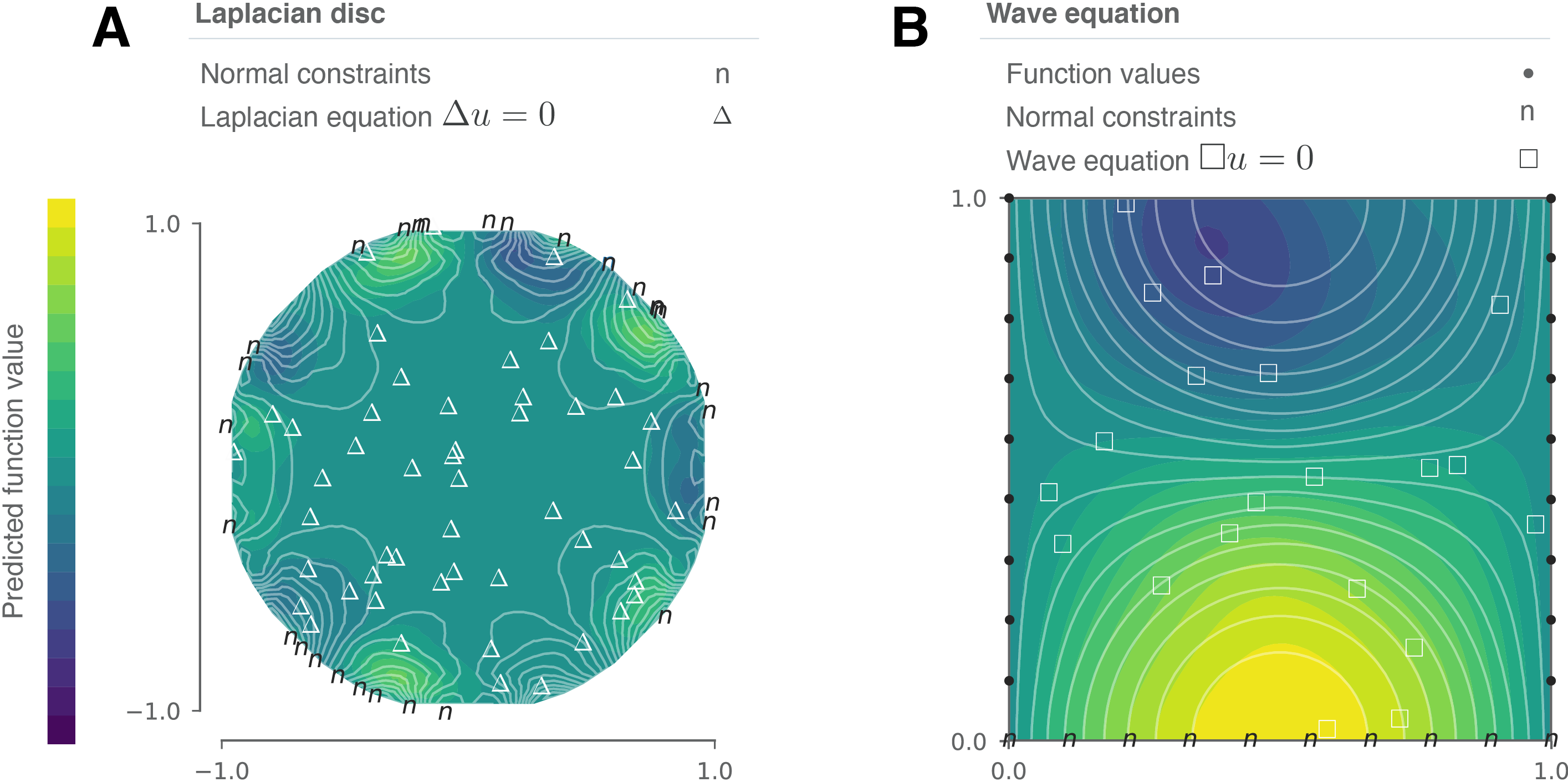}
    \caption{\textbf{Combining differential equation constraints.} (A) Solving Laplace's equation $\Delta u(x) = 0$ on the unit disk with Neumann boundary condition $\nabla u(x) \cdot \mathbf n(x)=\cos(5\phi)$ where $\phi$ is the radial angle of $x$, and $\mathbf n(x)$ is the boundary normal vector. The unknown function $u$ is modeled as a GP
    {(RBF kernel, length scale $\sigma = 0.5$)}
    The PDE is then solved using 1) Laplacian-transformed observations in the interior and 2) directional derivatives in the normal direction on the boundary as constraints. %
    (B) Using the same approach to solve a one-dimensional wave equation $\square u(x, t) = 0$, where $\square = \tfrac{\partial^2}{\partial t^2} - \tfrac{\partial^2}{\partial x^2}$, subject to Dirichlet boundary conditions $u(0,t)=u(1,t)=0$, $u(x,0)=x(1-x)$, and an initial Neumann boundary condition $\partial_t u(x, 0)=0$.
    }
    \label{fig:laplacegp}
\end{figure*}

An example were multiple DE constraints need to be combined within the same model is the Laplace equation on a two-dimensional disc with Neumann boundary conditions (see Fig.~\ref{fig:laplacegp}A):
\begin{align}
    \Delta u(\mathbf{x}) &= 0, \quad\quad\quad\,\,\,\, \mathbf{x} \in \Omega,\\
    \nabla u(\mathbf{x}) \cdot \mathbf n(\mathbf{x})&=\cos(5\phi), \quad \mathbf{x} \in \partial\Omega.
\end{align}
The combination of gradient and divergence constraints, including their cross-covariances, is united in the same GP kernel:
\begin{equation}
\mathbf{K}_{\text{LN}} = \left[\begin{array}{cc}
\mathbf{K}_{\Delta, \Delta}  & \mathbf{K}_{\Delta, \nabla}	\\
\mathbf{K}_{\nabla, \Delta} & \mathbf{K}_{\nabla, \nabla}	\\
\end{array}\right].
\end{equation}
In this setting, our small training dataset consists of divergence observations $\Delta u(\mathbf{x})$ in the interior as well as directional derivative observations $\nabla u(\mathbf{x}) \cdot \mathbf n(\mathbf{x})$ on the boundary. 
Here, we crucially rely on the extrapolation power due to the DE constraints to solve this reconstruction task with high accuracy.

In another example we add second boundary constraint type to solve a wave equation in one spatial dimension, $\mathbf{x}=(x,t)\in\Omega=[0,1]^2$ (see Fig.~\ref{fig:laplacegp}B). Here, the constraints of the d’Alembertian operator $\square = \tfrac{\partial^2}{\partial t^2} - \tfrac{\partial^2}{\partial x^2}$ are combined with Dirichlet (fixed function values) and Neumann conditions along the boundary of the domain, leading to a linear PDE:
\begin{align}
    \square u(x,t) &= 0, \quad\quad \mathbf{x} \in \Omega,\\
    u(0,t)&=0\\
    u(1,t)&=0\\
    u(x,0)&=x(1-x)\\
    \partial_t u(x,0)&=0.
\end{align}
We reiterate that the constrained kernels are not explicitly provided to any of the models in analytical form, but generated on-the-fly from a basic scalar-valued kernel function 
$k(\mathbf{x}, \mathbf{x}^{\prime})=\exp (-\tfrac{1}{2}\sigma^{-2}\left\|\mathbf{x}-\mathbf{x}^{\prime}\right\|^{2})$ 
(radial basis function (RBF) kernel) in all cases. We also remark that our model intentionally uses the RBF kernel and not the correct Green functions, e.g. the heat kernel for the Laplace equation. This mimics the typical ML scenario, were an accurate reconstruction is needed, but the underlying structure of the problem is not fully understood.

\section{Data and code availability}

We provide a reference implementation for all experiments at
\begin{center}
\begin{adjustbox}{max width=\columnwidth}
    \url{https://github.com/niklasschmitz/ad-kernels}
\end{adjustbox}
\end{center}
The MD17 data set is publicly available from \texttt{http://www.sgdml.org} \cite{chmiela2017}.

\begin{table*}[t]
    \centering
    \begin{adjustbox}{max width=\textwidth}
    \begin{tabular}{lcccccccc}
	\toprule
	                              & Aspirin        & Benzene        & Ethanol        & Malonaldehyde  & Naphthalene     & Salicylic acid & Toluene        & Uracil         \\
	\midrule
	    
	sGDML                         & 0.702          & 0.163          & 0.341          & 0.410          & 0.115          & 0.286          & 0.145          & 0.242          \\
	FCHL19                        & 0.628          & 0.179          & 0.180          & 0.302          & 0.185          & 0.277          & 0.246          & 0.147          \\
	\midrule 
	sGDML+FCHL19             & \textbf{0.430} & 0.158          & \textbf{0.160} & 0.274          & \textbf{0.094} & \textbf{0.172} & \textbf{0.111} & 0.130          \\
	sGDML+FCHL19[Mat\'{e}rn]      & 0.636    & 0.196          & 0.353          & 0.494          & 0.303          & 0.366          & 0.335          & 0.254          \\
	sGDML[RBF]+FCHL19[Mat\'{e}rn] & 0.645          & 0.198          & 0.356          & 0.500          & 0.312          & 0.377          & 0.347          & 0.265          \\
	sGDML[RBF]+FCHL19             & 0.484          & \textbf{0.157} & 0.168          & \textbf{0.292} & 0.082          & 0.178          & 0.116          & \textbf{0.127} \\
	\bottomrule
    \end{tabular}
    \end{adjustbox}
    \caption{\textbf{Constructing GPs using linear combinations of multiple kernels.} Test errors (MAE) for force learning on MD17 with 1000 reference points. All errors in kcal mol$^{-1}$ Å$^{-1}$. The best result for each dataset is marked in bold.} %
    \label{tab:remix_models2}
\end{table*}

\end{document}